\documentclass[article,twocolumn]{revtex4}
\usepackage{graphicx}
\usepackage{amssymb}
\usepackage{bm}
\usepackage{hyperref}
\usepackage{epstopdf}

\begin{document}\normalsize

\title{Finding Possibility of Dynamical Dark Energy with Hubble Parameters}

\author{Jiaxin Wang}
\email{jxw@mail.nankai.edu.cn}
\author{
Xinhe Meng \email{xhm@nankai.edu.cn}
}

\affiliation{
Department of Physics, Nankai University, Tianjin 300071, China\\
State Key Laboratory of Theoretical Physics, Institute of Theoretical Physics,
Chinese Academy of Science, Beijing 100190, China
}


\begin{abstract}
The Hubble parameter is a critical measurement in cosmology, which contains the most direct information of the cosmic expansion history. Since discrepancy is found between low redshift and high redshift estimations of Hubble constant, we are interested in whether that tension indicates dynamical dark energy. In this paper we emphasize that the observed Hubble parameters at various redshifts, along with observed Hubble constant, can help us in probing the evolutional behavior of the mysterious dark energy. Null hypothesis tests are carried out with two diagnostic approaches.
We find out that, according to the present measurements of Hubble parameters, rejection of constant dark energy is captured at $1\sigma$ level from null tests with and without the observed value of Hubble constant.
\end{abstract}

\maketitle

\section{Introduction}
The Constant Dark Energy Cold Dark Matter model ($\Lambda$CDM) --- which is more usually called the concordance model or the base model --- has  well fitted a variety of observations, proving that it is one of the most beautiful models in a mathematical view. Despite of its successfulness, we should admit that we still lack clear understanding of two main components in the Universe --- dark matter and dark energy. In order to gain more knowledge of the mysterious dark components, we have to constrain the parameters in proposed models with more observational data. We are quite confident that the cold dark matter model satisfies the requirements for the appearance of the large scale structure we observed at present, while we are not at the same level of confidence about whether the dark energy is truly a cosmological constant $\Lambda$. The useful results we had captured were almost all pointing to constant dark energy term with equation-of-state (EoS) parameter $w$ lies around $-1$. Lately, several researches have found indirect clues for dynamical dark energy from its reconstructed EoS~\cite{gbz} or from the discrepancy found between low-z (HST) and high-z (Planck) Hubble constant estimations~\cite{planck,hst}, which has been a hotspot in cosmology research.

We approach in a model independent way to investigate that tension via non-parametric methods (the original null test and improved one) to check the validity of the concordance model. Observed Hubble parameter Data (OHD (Tab.~\ref{tab:1})) which is different from SN and CMB measurements is adopted. Independent analysis with OHD alone or joint analysis of OHD and Hubble constant (HST~\cite{hst}) are both carried out.

The null hypothesis test --- the null test for short --- is a complementary test for the conventional model constraints with observational data~\cite{sahni}. Since null tests rely on data (and statistical methods if necessary) directly, they are useful in giving independent judgements about the validity of the proposed models. Previous researches focused on testing null hypothesis with OHD, BAO and SN data-sets, and also with different statistical approaches, ie., Gaussian processes (GP)~\cite{seikel} and goodness of fit (GoF) criterion combined with principle component analysis (PCA)~\cite{gof}. The observed Hubble constant was not usually emphasized in the previous work in probing dark energy, especially in null tests, since its function was usually considered as a normalization factor when converting $h(z)$ into the dimensionless Hubble parameter $E(z) = h(z)/h_0$. In this paper we define ${\rm H}(z) =100\cdot h(z)~km\cdot s^{-1}\cdot Mpc^{-1}$ and ${\rm H}_0 = 100\cdot h_0~km\cdot s^{-1}\cdot Mpc^{-1}$ for convenience, and estimation error of observed Hubble constant value will also be included. We also carried out two point diagnostic function~\cite{newtest} which enable us to conduct null tests with OHD independently.

According to the Refs.~\cite{seikel,yahya,gof}, the precision of a null test becomes lower when we require higher order of derivative value of data. Meanwhile, concerning about the quantity of the present OHD samples, the results of null tests under the assumption of flat space metric is more convincing than those under the assumption of curved space. We conduct both parameter constraints and null tests in the flat RW metric and with the same data-set, so the results provided by two approaches can be compared.

This paper is arranged as follows.
We list the samples of Hubble parameter data-set in section \ref{section I}, besides which conventional parameter constraints on $\Lambda$CDM is carried out with OHD independently. The null hypothesis tests using Hubble parameter with and without Hubble constant are conducted in section \ref{section II}. General discussions are given in the final section \ref{final}.

\section{samples} \label{section I}
In flat  RW metric, we can describe the $\Lambda$CDM as
\begin{eqnarray}
{E}^2(z)_{\Lambda CDM} &=& [\Omega_m(1+z)^3+(1-\Omega_m)],
\end{eqnarray}
where $\Omega_m$, the dimensionless matter density at present. We neglect the radiation component, since the ratio of radiation among the energy constituents is as small as about $10^{-5}$ at low redshift. Chi-square analysis is adopted for parameter constraints which is described as
\begin{equation}
\chi^2_h = [\frac{h(z)_{\rm obs}-h(z)_{\rm model}}{\sigma_{\rm obs}}]^2, \label{hz}
\end{equation}
where $h(z)_{\rm obs}$ and $\sigma_{\rm obs}$ are given by OHD listed in Tab.~\ref{tab:1}. The numerical methods adopted here are discussed in Ref.~\cite{press}.

Different research groups have given their observation results about the Hubble constant~\cite{planck,hst,ngc6264,ugc3789,chp}. In this section we adopt the value of $h_0$ from HST for the joint analysis of Hubble constant and Hubble parameters.

In terms of OHD, we have included the latest results of Hubble parameter measurement. The combined OHD set is listed in Tab.~\ref{tab:1}, where most of the observed data points
(\cite{h1,h2,h3,h4}) are given by ``cosmic chronometers'', which offers a measurement of the expansion rate without relying on the nature of metric between the chronometer and observer.
The Hubble parameters at different redshift are estimated independently from any cosmological models, by using age differences of old elliptical galaxies that are passively evolving~\cite{jimenez}.

The Hubble parameters estimated from galaxy surveys such as Wiggle-Z~\cite{h6} and SDSS~\cite{h7,h9,h5} are usually more or less affected by the concordance model (ie., using distance prior from CMB) assumed at the first place. In this paper, we include the most model independent data~\cite{h9,h5} where Hubble parameters are estimated without adopting a dark energy model or priors form CMB data.


\begin{table}[!ht]
\caption{Observational Hubble parameter Data set used in this paper.
}
\label{tab:1}
\begin{tabular}{cccc}
\hline\noalign{\smallskip}
    redshift &  $h(z)$ & $1\sigma$ error & Ref \\
\hline
0.100&	0.69&	0.12&	\cite{h1}\\
0.170&	0.83&	0.08&	\cite{h1}\\
0.270&	0.77&	0.14&	\cite{h1}\\
0.400&	0.95&	0.17&	\cite{h1}\\
0.900&	1.17&	0.23&	\cite{h1}\\
1.300&	1.68&	0.17&	\cite{h1}\\
1.430&	1.77&	0.18&	\cite{h1}\\
1.530&	1.40&	0.14&	\cite{h1}\\
1.750&	2.02&	0.40&	\cite{h1}\\
0.480&	0.97&	0.62&	\cite{h2}\\
0.880&	0.90&	0.40&	\cite{h2}\\
0.1791&	0.75&	0.04&	\cite{h3}\\
0.1993&	0.75&	0.05&	\cite{h3}\\
0.3519&	0.83&	0.14&	\cite{h3}\\
0.5929&	1.04&	0.13&	\cite{h3}\\
0.6797&	0.92&	0.08&	\cite{h3}\\
0.7812&	1.05&	0.12&	\cite{h3}\\
0.8754&	1.25&	0.17&	\cite{h3}\\
1.037&	154&	0.20&	\cite{h3}\\
0.07&	0.690&	0.196&		\cite{h4}\\
0.12&	0.686&	0.262&		\cite{h4}\\
0.20&	0.729&	0.296&		\cite{h4}\\
0.28&	0.888&	0.366&		\cite{h4}\\
\noalign{\smallskip}\hline
0.57	&       0.876	&   0.072& \cite{h9}\\
0.35&	0.813&   0.038&	\cite{h5}\\
\hline
\end{tabular}
\end{table}



\begin{figure}[h!]
\begin{center}
\includegraphics[width=0.35\textwidth]{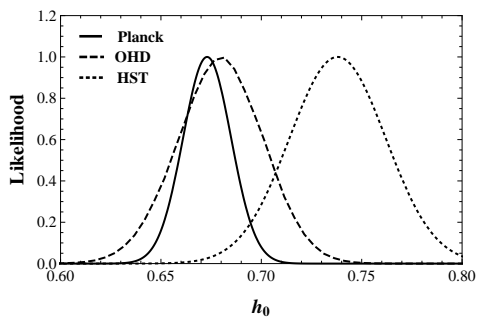}
\end{center}
\caption{Likelihood distribution of Hubble constant from HST, Planck and OHD separately, where the results given by Planck and OHD rely on the concordance model.}
\label{fig:00}
\end{figure}

For the concordance model, we capture the marginalized constraints on the Hubble constant given by OHD alone is $h_0 = 0.679\pm 0.021$ at $1\sigma$ level, which is consistent with HST's  and Planck's results within $1\sigma$, as shown in Fig.~\ref{fig:00}; while the result of HST ($h_0 = 0.738\pm 0.024$) is discrepant with Planck's estimation ($h_0 = 0.673\pm 0.012$) at around $2.5\sigma$ level~\cite{planck}, and the discrepancy between two measurements reaches $8.8\%$ and $9.7\%$ with respect to Planck and HST separately.

The results from parameter constraints are sensitive to the cosmological model assumed at the first place, especially the dimension and character of parameter spaces. In order to complete the analysis with observed Hubble constant and Hubble parameters, null hypothesis tests which are independent of any cosmological model are carried out in the following section.

\section{null tests} \label{section II}

\subsection{tests with smoothed data}

The general Friedmann equation with cold dark matter in flat RW metric reads
\begin{equation}
\frac{h(z)^2}{h_0^2} = \Omega_{m}(1+z)^3 + (1-\Omega_{m})\exp{[3\int_0^z \frac{1+w(z')}{1+z'}dz']} ,
\end{equation}
where $w(z')$, the equation-of-state (EoS) parameter of dark energy. Radiation is also neglected as we discussed above.

The EoS parameter of dark energy can be expressed as
\begin{equation}
w(z) = \frac{2(1+z)E E'-3E^2}{3[E^2-\Omega_m(1+z)^3]} ,\label{eos}
\end{equation}
where $E' = d E/d z$. Since we have no cosmological model independent measurement of parameter $\Omega_m$, the reconstruction of EoS of dark energy as Eq.~(\ref{eos}) from astrophysical data is not reliable. Null tests are  proposed as a solution which are expected to test the redshift dependency of dark energy with the null hypothesis that the dark energy is not dynamical~\cite{sahni}. The corresponding diagnostic function thus reads
\begin{equation}
\mathcal{O}_m = \frac{E^2-1}{z(3+3z+z^2)}. \label{o}
\end{equation}
The null hypothesis is true only if $\mathcal{O}_m$ remains as a constant in the redshift range covered by data.

Since we conduct the tests from observational data with estimation errors, the error of data will be transferred to Eq.~(\ref{o}), then the conclusion of the tests are weighted by their corresponding possibility (or the confidence level).
We should emphasize that the null-hypothesis test is different from the Bayesian analysis for constraining parameters. Following Fisher's framework~\cite{fisher}, the null hypothesis is potentially rejected or disproved on the basis of observational data rather than accepted. Thus in our work, we specify at which level the constant dark energy model is rejected.

The first order derivative of Eq.~(\ref{o}) which performs as an equivalent test but is more illustrative reads
\begin{equation}
\mathcal{L}_m = \frac{1}{(1+z)^6}[3(1+z)^2(1-E^2)+2z(3+3z+z^2)E E'], \label{l}
\end{equation}
where the factor $(1+z)^6$ is introduced in order to stabilise the errors at high redshift without affecting the consistency condition~\cite{seikel}.
The null hypothesis is rejected when the independently constructed Eq.~(\ref{l}) deviates from zero.
In the previous researches, different approaches have been carried out in order to seek robust estimation about the validity of the null hypothesis of dark energy.

In Eqs.~(\ref{o}) and~(\ref{l}), the observed value of Hubble constant --- acting as a normalization factor --- is necessary, the error and mean value of $h_0$ should be considered in order to giving convincing diagnoses.

We notice that the parameter space of $\Lambda$CDM model can be expressed by $\{ \Omega_m, h_0 \}$ or $\{\Omega_mh_0^2, h_0 \}$, thus in the parameter space $\{\Omega_mh_0^2, h_0 \}$ we can carry out null hypothesis with a diagnostic function which is equivalent to $\mathcal{O}_m$ as
\begin{equation}
\mathcal{D}_1 = \frac{h(z)^2-h_0^2}{z(3+3z+z^2)}, \label{h1}
\end{equation}
for emphasize the role of $h_0$ in null tests using observed Hubble constant.
The new diagnostic function can be reconstructed without normalizing $h(z)$ by $h_0$.

\begin{figure}[!]
\begin{center}
\includegraphics[width=0.35\textwidth]{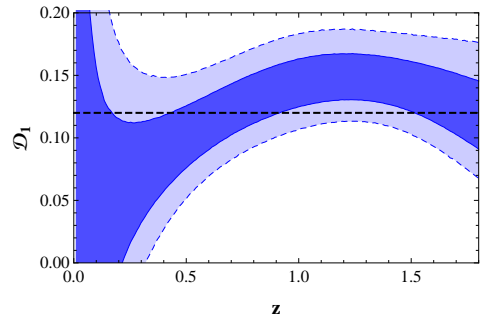}
\end{center}
\caption{Diagnostic function $\mathcal{D}_1$ with $1\sigma$ (dark blue) and $2\sigma$ (light blue) confidence limits of distribution given by HST+OHD, using GP method (which is introduced in Ref.~\cite{seikel}).}
\label{fig:smoothed}
\end{figure}

The result of $\mathcal{D}_1$ using GP method according to HST+OHD data-set is displayed in Fig.~\ref{fig:smoothed}, where deviation from  constant value is detected in the $1\sigma$ simulated region of $\mathcal{D}_1$, indicating that constant dark energy model is rejected at around $1\sigma$ level.

We should be careful about the strategy adopted for reconstructing $h(z)$ from  data, since different statistical methods may more or less lead to differences in smoothed curves of $h(z)$.
We think that analysis based on raw data seems to be more reliable for capturing convincing diagnoses.

\subsection{tests with raw data}

According to the utility of the Hubble constant in Eq.~(\ref{h1}), we can replace $h_0$ with the measured Hubble parameter value at the corresponding redshift point under the null hypothesis by
\begin{equation}
h(z_i)^2 = h_0^2[\Omega_m(1+z_i)^3+(1-\Omega_m)],
\end{equation}
where $z_i$ represents an arbitrary measurement point in the data-set (Tab.~\ref{tab:1}).
In the parameter space $\{\Omega_mh_0^2, h_0\}$, we can express $\Omega_mh_0^2$ as
\begin{equation}
\Omega_m h_0^2 = \frac{h(z)^2-h(z_i)^2}{(1+z)^3-(1+z_i)^3}.
\end{equation}
This expression suggests the improved diagnostic strategy free from smoothing $h(z)$ through observational data, which is originally introduced in Ref.~\cite{newtest} as the two point diagnostic function:
\begin{equation}
\mathcal{D}_{ij} =  \frac{h(z_j)^2-h(z_i)^2}{(1+z_j)^3-(1+z_i)^3}, \label{dij}
\end{equation}
where $i\neq j$, indicating two different OHD points.
Function~(\ref{h1}) can be recovered from $\mathcal{D}_{ij}$ when we choose $z_i = 0$, so this diagnostic function $\mathcal{D}_{ij}$ is a more general form which enable us to choose arbitrary reference points rather than $h_0$ only.

$\mathcal{D}_{ij}= \Omega_mh_0^2$ indicates the concordance model. If the new diagnostic function deviate from a constant, the validity of $\Lambda$CDM is in conflict with the present observations.
For each two different data points $\{ z_i, z_j \}$ in OHD, $\mathcal{D}_{ij}$ value can be estimated as a random variable with probability distribution simulated by observational data. We use the values of $1\sigma$ and $2\sigma$ limits to represent the distribution of each simulated value of diagnostic function according to the corresponding data.



\begin{figure}[!h]
\begin{center}
\includegraphics[width=0.35\textwidth]{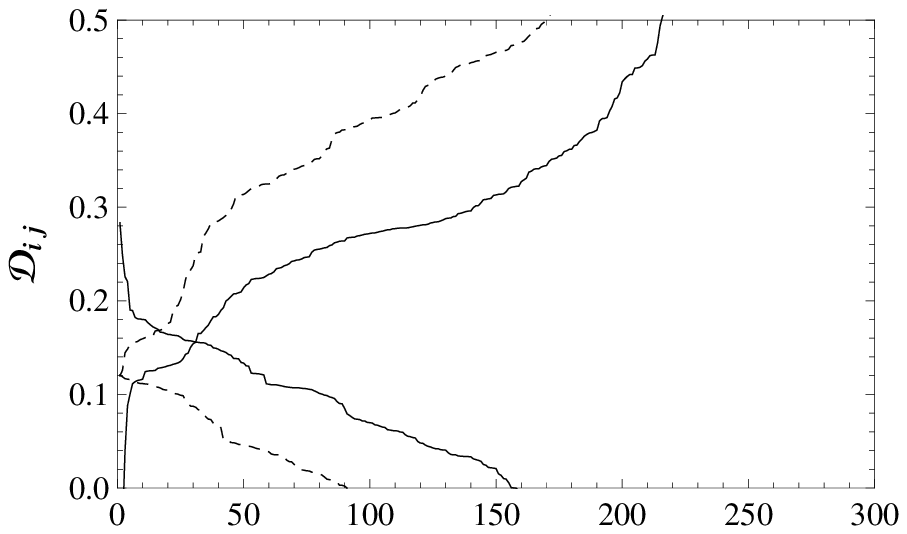}\\
\includegraphics[width=0.35\textwidth]{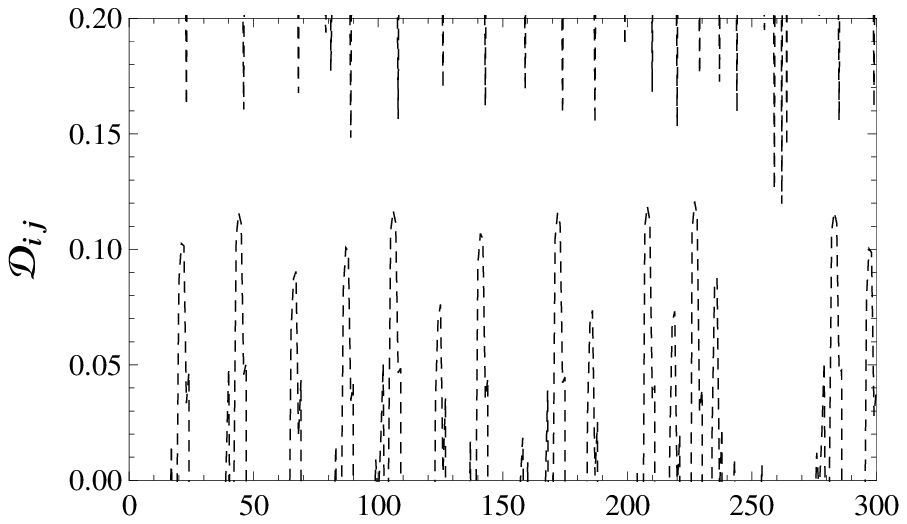}
\end{center}
\caption{{\it Upper panel}: The complete 300 independent results of Diagnostic function $\mathcal{D}_{ij}$ with $1\sigma$ (solid) and $2\sigma$ (dashed) limits generated from OHD. We rearranged the order of the points in each line so they are displayed in a nicer way. If null hypothesis is not deviated, then room for a constant should be allowed within the upper and lower limits. {\it Lower panel}: The $2\sigma$ limits of $\mathcal{D}_{ij}$ without being rearranged.}
\label{fig:7}
\end{figure}

According to OHD data-set, the full analysis of $\mathcal{D}_{ij}$ which contains $300$ independent results (which are extracted from the $25\times 25$ $\mathcal{D}_{ij}$ result matrix) is displayed in Fig.~\ref{fig:7} (with current OHD-set). Each upper and lower limit line consists of 300 independent values, the points in each upper (lower) limit line are rearranged from the lowest (highest) value to the highest (lowest) one. If the null hypothesis is supported by this test, an allowed region for constant value should exist from the full set of results, otherwise the lower and upper limit lines will cross with each other. We find that both $1\sigma$ and $2\sigma$ limit lines in the upper panel of Fig.~\ref{fig:7} cross each other. The $2\sigma$ deviation from null hypothesis is caused by the data at $z = 1.53$ as shown in the lower panel of Fig.~\ref{fig:7}, where the estimated values suddenly fall to about $0.11$ when $z_j = 1.53$. The $1\sigma$ rejection is not relieved even after we exclude the $1.53$ data point. This approach is much strict than previous null test methods and parameter constraints using the same data-set.

\begin{figure}[!h]
\begin{center}
\includegraphics[width=0.35\textwidth]{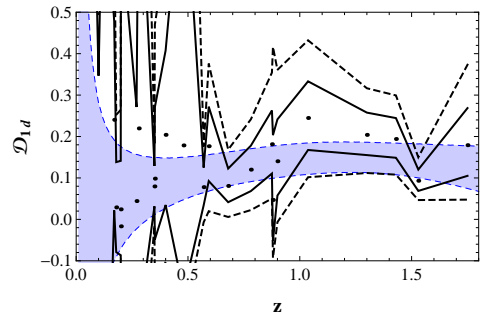}
\end{center}
\caption{the extra column characterized by $z_i = 0$ of $\mathcal{D}_{ij}$ (we name it as $\mathcal{D}_{1d}$), with $1\sigma$ (solid) and $2\sigma$ (dashed) limits. The dots represent the mean expectations and the colored background shows the $2\sigma$ range of $\mathcal{D}_1$.}
\label{fig:2}
\end{figure}

If we include observed value of $h$ in this null test, ie., the HST results, we will obtain an extra column of $\mathcal{D}_{ij}$. The extra results are shown in Fig.~\ref{fig:2}, where diagnoses with both smoothing method and raw data are carried out in order to provide comparisons.

The rejection of constant dark energy at $1\sigma$ level from the null test using $\mathcal{D}_{ij}$ and raw OHD data with and without HST is consistent with the other approaches carried out in this paper.
We should notice that in the current work, the rejection of $\Lambda$CDM captured by $\mathcal{D}_{ij}$ relying on the assumption that the standard error of each data has been well estimated and the error distributions are gaussian. Our conclusions drawn from the tests are based on the premise that we believe the adopted data is reliable.

\section{conclusions} \label{final}
In this paper we investigated the possibility of the existence of dynamical dark energy according to Hubble parameters, mainly through model independent null hypothesis test. We picked cosmological model independent data for the null hypothesis tests, and carried out two diagnostic methods~\cite{seikel,newtest} with OHD and HST in order to capture more robust results.

We adopted the improved the null test strategy through this work with two point diagnostic function~\cite{newtest}, which can be carried out without reconstruction method or relying on the estimation of Hubble constant as the only reference point. The results from the null tests show that constant dark energy is rejected at least at $1\sigma$ level according to OHD with or without HST.

Finding an ideal probing method at present is not easy, since each strategy has its own advantage and disadvantage. Parameter constraining method is sensitive to presumed models, while null tests which requiring reconstruction of $h(z)$ depend on the smoothing method. The two point diagnostic function seems to be able to offer better performance even with low quantity of data-set, but the conclusion is sensitive to the degree of confidence we have in each data point.

Since no strong deviation from concordance model is detected at $2\sigma$ level, more convincing evidences are required if we hope to conclude that $\Lambda$CDM model is inappropriate in describing the basic features of the Universe.
For better and more decisive conclusion about the truth of our universe, we are expecting more coming high quality observations at both high redshift and low redshift in order to conduct the diagnostic tests and parameter constraints in different redshift ranges.

\section*{Acknowledgements}
JW appreciates the helpful discussions and instructions from prof. Q.~Huang and Dr. C.~Cheng during his visit in ITP-CAS.


\newpage
\begin{thebibliography}{99}

\bibitem{gbz}
G.~Zhao, R.~G.~Crittenden, L.~Pogosian, X.~Zhang, Phys. Rev. Lett. {\bf 109} (2012) 171301.

\bibitem{planck}
Planck Collaboration, arXiv: 1303.5076.

\bibitem{hst}
A.~G.~Riess, {\it et al.}, Astrophys. J. {\bf 730} (2011) 119.

\bibitem{seikel}
M.~Seikel, {\it et al.}, Phys. Rev. D {\bf 86} (2012) 083001.

\bibitem{yahya}
S.~Yahya, M.~Seikel, C.~Clarkson, {\it et al.}, Phys. Rev. D {\bf 89} (2014) 023503.

\bibitem{gof}
Hao-Ran Yu, Shuo Yuan, Tong-Jie Zhang, Phys. Rev. D {\bf 88} (2013) 103528.

\bibitem{sahni}
V.~Sahni, A.~Shafieloo, A.~A.~Starobinsky, Phys. Rev. D {\bf 78} (2008) 103502.

\bibitem{fisher}
R.~A.~Fisher, {\it Statistical Mehtods for Research Workers}, Oliver \& Boyd, 1925.

\bibitem{newtest}
A.~Shafieloo, {\it et al.}, Phys. Rev. D {\bf 86} (2012) 103527.

\bibitem{press}
W.~H.~Press, {\it et al.}, {\it Numerical Recipes} 3rd Ed. Cambridge University Press, 2007.

\bibitem{ngc6264}
C.~Kuo, {\it et al.}, Astrophys. J. {\bf 767} ï¼?013ï¼?55.

\bibitem{ugc3789}
M.~J.~Reid, {\it et al.}, Astrophys. J. {\bf 767} (2013) 154.

\bibitem{chp}
W.~L.~Freedman, {\it et al.}, Astrophys. J. {\bf 758} (2012) 24.

\bibitem{h1}
J.~Simon, L.~Verde, R.~Jimenez, Phys. Rev. D {\bf 71} (2005) 123001.

\bibitem{h2}
D.~Stern, R.~Jimenez, L.~Verde, M.~Kamionkowski, S.~A.~Standford, JCAP {\bf 02} (2010) 008.

\bibitem{h3}
M.~Moresco, A.~Cimatti, R.~Jimenez, {\it et al.}, JCAP {\bf 1208} (2012) 006.

\bibitem{h4}
Cong Zhang, Han Zhang, Shuo Yuan, Tong-Jie Zhang, Yan-Chun Sun, arXiv:1207.4541.

\bibitem{h6}
C.~Blake, S.~Brough, M.~Colless, {\it et al.}, Mon. Not. Roy. Astron. Soc. {\bf 425} (2012) 405.

\bibitem{h7}
N.~G.~Busca, {\it et al.}, Astron. Astro. {\bf 552} (2013) A96.

\bibitem{h9}
C. Chuang, {\it et al.}, arXiv:1303.4486.

\bibitem{h5}
M.~Hemantha, Y. Wang, C.~Chuang, arXiv:1310.6468.

\bibitem{h8}
A. Font-Ribera, {\it et al.}, arXiv:1311.1767.

\bibitem{jimenez}
R.~Jimenez, A.~Loeb, Astrophys J. {\bf 573} (2002) 37.

\bibitem{code}
M.~Seikel, C.~Clarkson, M.~Smith, JCAP {\bf 1206} (2012) 036.


\end{thebibliography}
\end{document}